\documentclass{article}
\usepackage{amsmath,amssymb,color,url,amsthm}
\usepackage[pdftex,hiresbb]{graphicx}
\usepackage{type1cm}
\usepackage{subcaption}
\usepackage{geometry}
\allowdisplaybreaks
\newtheorem{thm}{Theorem}[section]

\newtheorem{lem}[thm]{Lemma}

\newtheorem{rem}[thm]{Remark}
\newtheorem{ass}[thm]{Assumption}
\newtheorem{ex}[thm]{Example}
\def\l     {\left}
\def\r     {\right}
\def\<     {\langle}
\def\>     {\rangle}
\def\fin   {\hfill{$\Box$}\vspace{5mm}}
\def\bbE   {{\mathbb E}}
\def\bbP   {{\mathbb P}}
\def\bbR   {{\mathbb R}}

\def\calD  {{\mathcal D}}
\def\calF  {{\mathcal F}}
\def\calL  {{\mathcal L}}
\def\calV  {{\mathcal V}}
\def\olcalL{{\overline \calL}}
\def\vt    {\vartheta}

\def\tN    {\widetilde{N}}

\def\whS   {\widehat{S}}

\begin{document}
\title{Al\`os type decomposition formula for Barndorff-Nielsen and Shephard model}
\author{Takuji Arai\footnote{Department of Economics, Keio University, 2-15-45 Mita, Minato-ku, Tokyo, 108-8345, Japan. (arai@econ.keio.ac.jp)}}
\maketitle

\begin{abstract}
The objective is to provide an Al\`os type decomposition formula of call option prices for the Barndorff-Nielsen and Shephard model:
an Ornstein-Uhlenbeck type stochastic volatility model driven by a subordinator without drift.
Al\`os \cite{Alos12} introduced a decomposition expression for the Heston model by using Ito's formula.
In this paper, we extend it to the Barndorff-Nielsen and Shephard model.
As far as we know, this is the first result on the Al\`os type decomposition formula for models with infinite active jumps. 
\end{abstract}

%
%
\setcounter{equation}{0}
\section{Introduction}
Stochastic volatility models have drawn considerable attention in mathematical finance since they are very useful for capturing the volatility skew and smiles,
but there is no closed-form option pricing formula for stochastic volatility models in general.
Thus, some authors have presented decomposition expressions of option prices, which are useful to derive approximations of option prices and to analyze implied volatilities.
Firstly, for continuous stochastic volatility models with no correlation between the asset price and the volatility processes,
Hull and White \cite{HW87} provided an option price expression with a conditional expectation of the Black-Scholes formula
by substituting the future average volatility for the volatility in the Black-Scholes formula.
Al\`os \cite{Alos06} has extended it to correlated models by means of Malliavin calculus in order to deal with Ito's formula for anticipating processes,
since the future average volatility is a non-adapted process.
Besides, extensions to more general models have been done by \cite{ALPV09}, \cite{ALV07}, \cite{JV13} and so on.
On the other hand, Al\`os \cite{Alos12} obtained a new decomposition formula for the Heston model by using the average squared future volatility, instead of the future average volatility.
Since the average squared future volatility is an adapted process, she made use of the classical Ito calculus, not the Malliavin calculus.
The decomposition formula in \cite{Alos12} is given as the sum of the Black-Scholes formula and terms due to the volatility process.
In addition, using the obtained decomposition expression, approximate option pricing formulas were also presented.
This Al\`os type decomposition formula has been extended to more general models by \cite{MV15}, \cite{MV17} and so on.
Among them, Merino et al. \cite{MPSV18} has extended to stochastic volatility models with finite active jumps.
Moreover, for the Heston model, Al\`os et al. \cite{ADV15} suggested an approximation of the implied volatility and a calibration method by using the results of \cite{Alos12}.

The objective of this paper is to obtain an Al\`os type decomposition expression of call option prices for the Barndorff-Nielsen and Shephard (BNS) model
by applying Ito's formula to the Black-Scholes formula.
It is given as the sum of the Black-Scholes formula, a term due to the impact of the asset price jumps, and some residual terms due to the asset price jumps and changes of the volatility.
Unlike \cite{Alos12}, we use the current squared volatility value instead of the average squared future volatility, and substitute it to the volatility in the Black-Scholes formula.
To our best knowledge, this is the first result of the Al\`os type decomposition formula for models with infinite active jumps,
but Jafari and Vives \cite{JV13} derived a Hull-White type decomposition formula for models with infinite active jumps by means of Mailliavin calculus.
Now, the BNS model is a representative jump-type stochastic volatility model undertaken by \cite{BNS1}, \cite{BNS2},
and its volatility process is given by a non-Gaussian Ornstein-Uhlenbeck process.
For details on the BNS model, see also \cite{NV} and \cite{Scho}.
The BNS model has the following three features:
First, the asset price process has jumps, but all jumps are negative.
Second, there is no Brownian component in the volatility process.
Third, the jump component is common between the asset price and the volatility processes.
Remark that the jumps might be infinite active.
Our decomposition formula will be derived by making the most of these features of the BNS model.

The structure of this paper is as follows: We give some mathematical preliminaries and notations in the following section.
Section 3 introduces our decomposition formula.
Its proof is given in Section 4, and conclusions are summarized in Section 5.

%
%
\setcounter{equation}{0}
\section{Preliminaries}
\subsection{Model description}
Consider throughout a financial market model in which only one risky asset and one riskless asset are tradable.
Let $r\geq0$ be the interest rate of our market, and $T>0$ a finite time horizon.
In the BNS model, the risky asset price at time $t\in[0,T]$ is described by
\begin{equation}\label{eq-S}
S_t:=S_0\exp\l\{\int_0^t\l(r+\mu-\frac{1}{2}\Sigma_u^2\r)du+\int_0^t\Sigma_udW_u+\rho H_{\lambda t}\r\}, \ \ \ t\in[0,T],
\end{equation}
where $S_0>0$, $\rho\leq0$, $\mu\in\bbR$, $\lambda>0$, $H$ is a subordinator without drift, and $W$ is a $1$-dimensional standard Brownian motion.
Here $\Sigma$ is the volatility process, of which squared process $\Sigma^2$ is given by an Ornstein-Uhlenbeck process driven by the subordinator $H_\lambda$, that is,
the solution to the following stochastic differential equation:
\begin{equation}\label{SDE-sigma}
d\Sigma_t^2 = -\lambda\Sigma_t^2dt+dH_{\lambda t}, \ \ \ t\in[0,T]
\end{equation}
with $\Sigma_0^2>0$.
Note that the asset price process $S$ is defined on some filtered probability space $(\Omega,\calF,(\calF_t)_{0\leq t\leq T},\bbP)$ with the usual condition,
where $(\calF_t)_{0\leq t\leq T}$ is the filtration generated by $W$ and $H_\lambda$.
In addition, we denote by $X$ the log price process $\log S$, that is,
\begin{equation}\label{def-X}
X_t:=\log S_t = \log S_0+\int_0^t\l(r+\mu-\frac{1}{2}\Sigma_u^2\r)du+\int_0^t\Sigma_udW_u+\rho H_{\lambda t}, \ \ \ t\in[0,T].
\end{equation}
Remark that the term $\rho H_{\lambda t}$ in (\ref{def-X}) (or (\ref{eq-S})) accounts for the leverage effect,
which is a stylized fact such that the asset price declines at the moment when the volatility increases.

For later use, we enumerate some properties of $\Sigma$: Firstly, we have
\begin{equation}\label{sigma1}
\Sigma_t^2 = e^{-\lambda t}\Sigma^2_0+\int_0^te^{-\lambda(t-u)}dH_{\lambda u} \geq e^{-\lambda T}\Sigma^2_0
\end{equation}
for any $t\in[0,T]$, that is, $\Sigma$ is bounded from below.
Next, the integrated squared volatility is represented as
\begin{equation}\label{sigma2}
\int_t^T\Sigma_u^2du=\epsilon(T-t)\Sigma^2_t+\int_t^T\epsilon(T-u)dH_{\lambda u}
\end{equation}
for any $t\in[0,T]$, where
\[
\epsilon(t):=\frac{1-e^{-\lambda t}}{\lambda}.
\]
In addition, (\ref{sigma2}) implies
\begin{equation}\label{sigma3}
\int_0^T\Sigma_u^2du\leq \frac{1}{\lambda}(H_{\lambda T}+\Sigma^2_0).
\end{equation}

Now, we denote by $N$ the Poisson random measure of $H_\lambda$.
Hence, we have
\[
H_{\lambda t}=\int_0^\infty zN([0,t],dz), \ \ \ t\in[0,T].
\]
Letting $\nu$ be the L\'evy measure of $H_\lambda$, we find that
\[
\tN(dt,dz):=N(dt,dz)-\nu(dz)dt
\]
is the compensated Poisson random measure.
Note that $\nu$ is a $\sigma$-finite measure on $(0,\infty)$ satisfying
\[
\int_0^\infty(z\wedge1)\nu(dz)<\infty
\]
by Proposition 3.10 of \cite{CT}.
The asset price process $S$ is also given as the solution to the following stochastic differential equation:
\[
dS_t=S_{t-}\l\{\alpha dt+\Sigma_tdW_t+\int_0^\infty(e^{\rho z}-1)\tN(dt,dz)\r\}, \ \ \ t\in[0,T],
\]
where
\[
\alpha:=r+\mu+\int_0^\infty(e^{\rho z}-1)\nu(dz).
\]
Note that $S_t>0$ holds for any $t\in[0,T]$.

Now, we introduce our standing assumption as follows:

\begin{ass}\label{ass0}
\begin{enumerate}
\item $\displaystyle{\mu=\int_0^\infty(1-e^{\rho z})\nu(dz)}$.
\item $\displaystyle{\int_1^\infty e^{2\epsilon(T)z}\nu(dz)<\infty}$.
\end{enumerate}
\end{ass}

\noindent
The above condition 1 implies that the discounted asset price process $\whS_t:=e^{-rt}S_t$ becomes a local martingale.
On the other hand, the condition 2 ensures that
\[
\int_0^\infty z^2\nu(dz)<\infty,
\]
which yields $\bbE[H_{\lambda T}^2]<\infty$ by Proposition 3.13 of \cite{CT}, and
\begin{equation}\label{eq-X2}
\bbE\l[\sup_{t\in[0,T]}X^2_t\r]<\infty
\end{equation}
by (\ref{sigma3}).
In addition,
\begin{equation}\label{eq-S2}
\bbE\l[\sup_{t\in[0,T]}S^2_t\r]<\infty
\end{equation}
holds under the condition 2 from the view of Subsection 2.3 of \cite{AS}.
Thus, $\whS$ is a square-integrable martingale under Assumption \ref{ass0}.

\begin{ex}\label{ex1}
We introduce two important examples of the squared volatility process $\Sigma^2$.
\begin{enumerate}
\item The first one is the case where $\Sigma^2$ follows an IG-OU process. The corresponding L\'evy measure $\nu$ is given by
      \[
      \nu(dz)=\frac{\lambda a}{2\sqrt{2\pi}}z^{-\frac{3}{2}}(1+b^2z)\exp\l\{-\frac{1}{2}b^2z\r\}dz, \ \ \ z\in(0,\infty),
      \]
      where $a>0$ and $b>0$.
      Note that this is a representative example of the BNS model with infinite active jumps, that is, $\nu((0,\infty))=\infty$.
      In this case, the invariant distribution of $\Sigma^2$ follows an inverse-Gaussian distribution with parameters $a>0$ and $b>0$.
      Note that the condition 2 of Assumption \ref{ass0} is satisfied whenever $\frac{b^2}{2}>2\epsilon(T)$
\item The second example is the gamma-OU case. In this case, $\nu$ is described as
      \[
      \nu(dz)=\lambda abe^{-bz}dz, \ \ \ z\in(0,\infty),
      \]
      and the invariant distribution of $\Sigma^2$ is given by a gamma distribution with parameters $a>0$ and $b>0$.
      If $b>2\epsilon(T)$, then the condition 2 of Assumption \ref{ass0} is satisfied.
\end{enumerate}
\end{ex}

\subsection{Black-Scholes formula}
In this subsection, consider the so-called Black-Scholes model with volatility $\sigma>0$ and interest rate $r\geq0$, and the call option with strike price $K>0$ and maturity $T>0$.
We describe the call option price at time $t\in[0,T)$ with the log asset price $x\in\bbR$ by a function $BS$ on not only $t$ and $x$, but also squared volatility $\sigma^2$.
Thus, the function $BS(t,x,\sigma^2)$, which is well-known as the Black-Scholes formula, is given as
\begin{equation}\label{def-BS}
BS(t,x,\sigma^2):=e^x\Phi(d^+)-Ke^{-r\tau_t}\Phi(d^-), \ \ \ t\in[0,T), x\in\bbR, \sigma>0,
\end{equation}
where $\tau_t = T-t$, $\Phi$ is the cumulative distribution function of the standard normal distribution, and
\begin{equation}\label{def-d1}
d^\pm:=\frac{x-\log K+r\tau_t}{\sigma\sqrt{\tau_t}}\pm\frac{\sigma\sqrt{\tau_t}}{2}.
\end{equation}
For later use, we denote
\begin{equation}\label{eq-BS1}
x_z := x+\rho z, \ \sigma_z := \sqrt{\sigma^2+z}, \ \eta^\pm := r\pm\frac{\sigma^2}{2}, \ \eta^\pm_z := r\pm\frac{\sigma_z^2}{2} = \eta^\pm\pm\frac{z}{2}
\end{equation}
for $z>0$, $x\in\bbR$ and $\sigma>0$.
Thus, $d^\pm$ is rewritten as
\[
d^\pm=\frac{x-\log K+\eta^\pm\tau_t}{\sigma\sqrt{\tau_t}}.
\]
Furthermore, we define
\begin{equation}\label{def-d2}
d^\pm_{\rho z}:=\frac{x_z-\log K+\eta^\pm\tau_t}{\sigma\sqrt{\tau_t}}=d^\pm+\frac{\rho z}{\sigma\sqrt{\tau_t}}.
\end{equation}
and
\begin{equation}\label{def-d3}
d^\pm_{\rho z,z}:=\frac{x_z-\log K+\eta^\pm_z\tau_t}{\sigma_z\sqrt{\tau_t}}
\end{equation}
for $z>0$.
Remark that the time parameter $t$ included in $d^\pm$, $d^\pm_{\rho z}$ and $d^\pm_{\rho z,z}$ might be replaced with $u$ or $s$ according to the situation.
In addition, since we have
\[
\lim_{t\to T}BS(t,x,\sigma^2)=(e^x-K)^+,
\]
the domain of the function $BS$ can be extended to $[0,T]\times\bbR\times(0,\infty)$, and we may define
\[
BS(T,x,\sigma^2):=(e^x-K)^+.
\]
For simplicity, substituting $X_t$ and $\Sigma_t^2$ defined in (\ref{def-X}) and (\ref{SDE-sigma}) for $x$ and $\sigma^2$ respectively in the function $BS$,
we denote
\[
BS_t:=BS(t,X_t,\Sigma^2_t)
\]
for $t\in[0,T]$.

More importantly, defining an operator $\calD^{BS}$ as
\[
\calD^{BS}f(t,x,\sigma^2):=\l(\partial_t+\frac{\sigma^2}{2}\partial^2_x+\eta^-\partial_x-r\r)f(t,x,\sigma^2)
\]
for $\bbR$-valued function $f(t,x,\sigma^2)$, $t\in[0,T)$, $x\in\bbR$, $\sigma>0$, we have
\begin{equation}\label{BS-PDE}
\calD^{BS}BS(t,x,\sigma^2)=0, \ \ \ t\in[0,T), x\in\bbR, \sigma>0.
\end{equation}
Remark that partial derivatives of $BS$ are given as
\begin{equation}\label{eq-dBSx}
\partial_xBS(t,x,\sigma^2)=e^x\Phi(d^+),
\end{equation}
\begin{equation}\label{eq-dBSx2}
\partial^2_xBS(t,x,\sigma^2)=e^x\Phi(d^+)+\frac{e^x}{\sigma\sqrt{\tau_t}}\phi(d^+),
\end{equation}
and
\begin{equation}\label{eq-G}
\partial_{\sigma^2}BS(t,x,\sigma^2)=\frac{\tau_t}{2}(\partial^2_x-\partial_x)BS(t,x,\sigma^2)=\frac{\sqrt{\tau_t}}{2\sigma}e^x\phi(d^+),
\end{equation}
where $\phi$ is the probability density function of the standard normal distribution.
All of the above derivatives are positive functions.
For later use, we define additionally the following operators for $\bbR$-valued function $f(t,x,\sigma^2)$, $t\in[0,T)$, $x\in\bbR$, $\sigma>0$:
\[
\Delta^{a,b}f(t,x,\sigma^2):=f(t,x+a,\sigma^2+b)-f(t,x,\sigma^2), \ \ \ a,b\in\bbR,
\]
\[
\calL^zf(t,x,\sigma^2):=\Delta^{\rho z,0}f(t,x,\sigma^2)+\partial_xf(t,x,\sigma^2)(1-e^{\rho z}), \ \ \ z>0,
\]
and
\[
\olcalL f(t,x,\sigma^2):=\int_0^\infty\calL^zf(t,x,\sigma^2)\nu(dz).
\]

%
%
\setcounter{equation}{0}
\section{Main results}
In this section, we introduce our main result, that is, a decomposition formula for the BNS model introduced in Section 2.
Recall that the discounted asset price process $\whS$ is a square-integrable martingale under Assumption \ref{ass0}.
Thus, for the vanilla call option with strike price $K>0$ and maturity $T>0$, its price at time $t\in[0,T]$ is given as
\[
V_t:=e^{-r\tau_t}\bbE[BS_T|X_t,\Sigma^2_t].
\]
In Theorem \ref{thm1} below, we derive a decomposition expression of $V_t$ by applying Ito's formula to the Black-Scholes function $BS$.
Its proof is postponed until Section 4.

\begin{thm}\label{thm1}
Under Assumption \ref{ass0}, we have, for $t\in[0,T]$,
\begin{equation}\label{eq-thm1-0}
V_t = BS_t+\tau_t\olcalL BS_t+I_1+I_2+I_3+I_4+I_5.
\end{equation}
Here, $I_1,\dots,I_5$ are defined as follows:
\[
I_1:=\bbE\l[\int_t^Te^{-r(u-t)}\partial_{\sigma^2}BS_u(-\lambda\Sigma^2_u)du\Big|X_t,\Sigma^2_t\r],
\]
\[
I_2:=\bbE\l[\int_t^Te^{-r(u-t)}\int_0^\infty\l(\Delta^{\rho z,z}-\Delta^{\rho z,0}\r)BS_u\nu(dz)du\Big|X_t,\Sigma^2_t\r],
\]
\[
I_3:=\bbE\l[\int_t^Te^{-r(u-t)}\tau_u\partial_x\olcalL BS_u\mu du\Big|X_t,\Sigma^2_t\r],
\]
\[
I_4:=\bbE\l[\int_t^Te^{-r(u-t)}\tau_u\partial_{\sigma^2}\olcalL BS_u(-\lambda\Sigma^2_u)du\Big|X_t,\Sigma^2_t\r],
\]
and
\[
I_5:=\bbE\Bigg[\int_t^Te^{-r(u-t)}\tau_u\int_0^\infty\Delta^{\rho z,z}\olcalL BS_u\nu(dz)du\Big|X_t,\Sigma^2_t\Bigg],
\]
where $\tau_u:=T-u$.
\end{thm}

\begin{rem}\label{rem1}
In the decomposition formula (\ref{eq-thm1-0}), the first two terms in the right hand side are regarded as principal terms.
In particular, the second term $\tau_t\olcalL BS_t$ represents the impact of the jumps of the asset price process. Indeed, it becomes 0 whenever $\rho=0$.
Note that this term converges to 0 with order 1 as the time to maturity $\tau_t$ tends to 0.
Here we give interpretations of $I_1,\dots,I_5$ in turn.
First of all, we can say that $I_1$ represents the influence of the continuous fluctuation of the squared volatility process $\Sigma^2$.
Next, decomposing $I_2$ into the following two terms
\begin{equation}\label{eq-rem1-1}
\bbE\l[\int_t^Te^{-r(u-t)}\int_0^\infty\Delta^{0,z}BS_u\nu(dz)du\Big|X_t,\Sigma^2_t\r],
\end{equation}
and
\begin{equation}\label{eq-rem1-2}
\bbE\l[\int_t^Te^{-r(u-t)}\int_0^\infty\l(\Delta^{\rho z,z}-\Delta^{\rho z,0}-\Delta^{0,z}\r)BS_u\nu(dz)du\Big|X_t,\Sigma^2_t\r],
\end{equation}
we can say that (\ref{eq-rem1-1}) represents the impact of the jumps of the squared volatility process,
but (\ref{eq-rem1-2}) is corresponding to the impact of that jumps occur simultaneously in the asset price process and the squared volatility process.
As for the last three terms, the comparison between (\ref{eq-thm1-0}) and (\ref{eq-thm1-3}) below gives
\[
I_3+I_4+I_5=\bbE\l[\int_t^Te^{-r(u-t)}\olcalL BS_udu\Big|X_t,\Sigma^2_t\r]-\tau_t\olcalL BS_t.
\]
Thus, the sum $I_3+I_4+I_5$ is corresponding to the residual part of the impact of the asset price jumps.
Each $I_3$, $I_4$ and $I_5$ represents the interaction of the impact of the asset price jumps with the continuous fluctuation of the asset price process,
the continuous fluctuation of the squared volatility process, and the fact that jumps occur simultaneously in the asset price and the squared volatility processes, respectively.
\end{rem}

\begin{rem}\label{rem2}
As mentioned in Section 1, the decomposition formula (\ref{eq-thm1-0}) is given as an extension of the result of \cite{Alos12} for Heston model,
in which the average squared future volatility $\calV^2_t$ has been substituted for the volatility in the Black-Scholes formula, where $\calV^2_t$ is defined as
\[
\calV^2_t:=\frac{1}{\tau_t}\int_t^T\bbE[\Sigma^2_u|\Sigma^2_t]du.
\]
Note that $\calV^2_t$ for the BNS model is given as 
\[
\calV^2_t=\frac{\epsilon(\tau_t)}{\tau_t}\Sigma^2_t+\frac{1}{\lambda}\l(1-\frac{\epsilon(\tau_t)}{\tau_t}\r)\int_0^\infty z\nu(dz)
\]
by (\ref{sigma1}).
In this paper, we use the current squared volatility value $\Sigma^2_t$, not $\calV^2_t$, since the use of $\Sigma^2_t$ simplifies our calculations drastically.
In addition, as indicated in Figure \ref{fig1} below, the difference between $BS_t=BS(t,X_t,\Sigma^2_t)$ and $BS(t,X_t,\calV^2_t)$ is sufficiently small.
Thus, the choice of $\Sigma^2_t$ or $\calV^2_t$ does not make a big impact.
\end{rem}

\setcounter{figure}{0}
\renewcommand{\figurename}{Panel}
\renewcommand{\thefigure}{(\alph{figure})}
\begin{figure}[htbp]
 \begin{minipage}{0.5\hsize}
    \includegraphics[width=70mm]{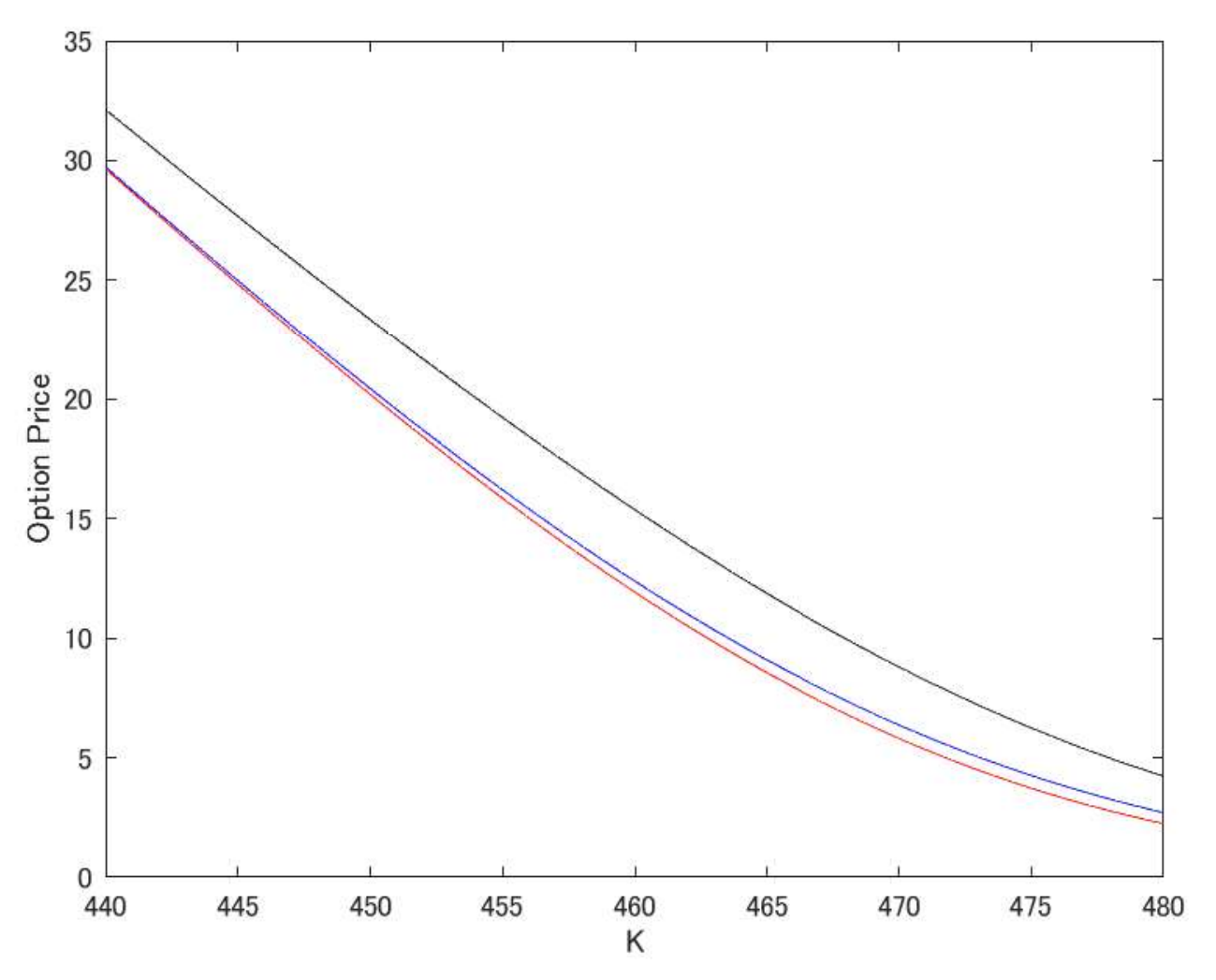}\vspace{-2mm}\caption{}\label{fig1a}
 \end{minipage}
 \begin{minipage}{0.5\hsize}
    \includegraphics[width=70mm]{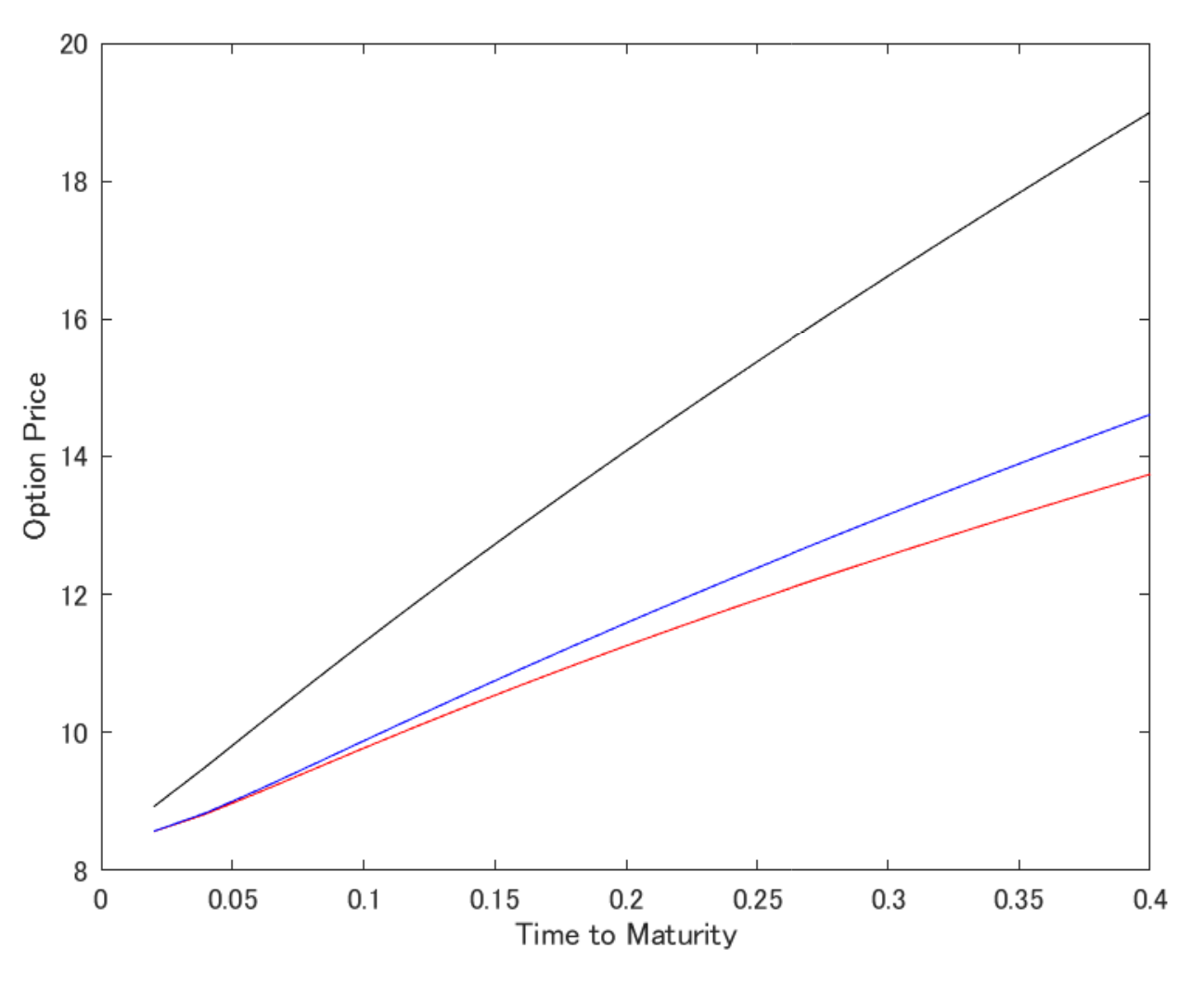}\vspace{-2mm}\caption{}\label{fig1b}
 \end{minipage}
\renewcommand{\figurename}{Figure}\setcounter{figure}{0}
\renewcommand{\thefigure}{\arabic{figure}}
\vspace{-2mm}\caption{We consider the IG-OU case of the BNS model introduced in Example \ref{ex1}.
We fix $t$ to 0 and set $\rho=-4.7039$, $\lambda=2.4958$, $a=0.0872$, $b=11.98$, $r=0.01$, $S_0=468.44$ and $\Sigma_0=0.064262$,
where this parameter set comes from Table 5.1 of \cite{NV}, who used S\&P 500 index option price data on November 2, 1993.
Remark that the above parameter set meets Assumption \ref{ass0}.
In this figure, we compute the values of $V_0$, $BS(0,X_0,\Sigma^2_0)$ and $BS(0,X_0,\calV^2_0)$.
Note that the values of $V_0$ are computed by the fast Fourier transform-based numerical scheme developed in Section 6 of \cite{AIS-BNS}
in order to compute the local risk-minimizing strategies for the BNS model as an extension of the so-called Carr-Madan method.
Panel \ref{fig1a} shows the values of $V_0$, $BS(0,X_0,\Sigma^2_0)$ and $BS(0,X_0,\calV^2_0)$ for the call options with strike price $K=440,440.1,\dots,480$
when the maturity $T$ is fixed to 0.25.
In Panel \ref{fig1b}, fixing $K$ to 460, and moving $T$ instead from 0.02 to 0.40 at steps of 0.02, we compute the same values for the option with $K=460$.
The black, red and blue curves represent the values of $V_0$, $BS(0,X_0,\Sigma^2_0)$ and $BS(0,X_0,\calV^2_0)$, respectively.}\label{fig1}
\end{figure}

\newpage
%
%
\setcounter{equation}{0}
\section{Proof of Theorem \ref{thm1}}
We shall show Theorem \ref{thm1} by applying Ito's formula twice to the Black-Scholes function.

\noindent{\it Step 1.} \
Fix $s,t\in[0,T)$ with $s>t$ arbitrarily for the time being.
Note that the function $e^{-ru}BS_u$, $u\in[s,t]$ is sufficiently smooth to apply Ito's formula.
From the view of Lemma \ref{lem2} below, we have
\begin{align}\label{eq-thm1-1}
e^{-rs}BS_s &= e^{-rt}BS_t-r\int_t^se^{-ru}BS_udu \nonumber \\
            &\hspace{5mm}+\int_t^se^{-ru}\partial_tBS_udu+\int_t^se^{-ru}\partial_xBS_u\l(r+\mu-\frac{\Sigma^2_u}{2}\r)du \nonumber \\
            &\hspace{5mm}+\int_t^se^{-ru}(\partial_xBS_u)\Sigma_udW_u+\frac{1}{2}\int_t^se^{-ru}(\partial^2_xBS_u)\Sigma^2_udu \nonumber \\
            &\hspace{5mm}+\int_t^se^{-ru}\partial_{\sigma^2}BS_u(-\lambda\Sigma^2_u)du \nonumber \\
            &\hspace{5mm}+\int_t^se^{-ru}\int_0^\infty\Delta^{\rho z,z}BS_{u-}N(du,dz) \nonumber \\
            &= e^{-rt}BS_t+\int_t^se^{-ru}\calD^{BS}BS_udu+\int_t^se^{-ru}\partial_xBS_u\mu du \nonumber \\
            &\hspace{5mm}+\int_t^se^{-ru}(\partial_xBS_u)\Sigma_udW_u+\int_t^se^{-ru}\partial_{\sigma^2}BS_u(-\lambda\Sigma^2_u)du \nonumber \\
            &\hspace{5mm}+\int_t^se^{-ru}\int_0^\infty\Delta^{\rho z,z}BS_{u-}N(du,dz).
\end{align}

Now, we take the conditional expectation given $X_t$ and $\Sigma^2_t$ on both sides of (\ref{eq-thm1-1}).
By (\ref{BS-PDE}) and Lemmas \ref{lem1} and \ref{lem2}, we have
\begin{align}\label{eq-thm1-2}
e^{-rs}\bbE[BS_s|X_t,\Sigma^2_t] &= e^{-rt}BS_t+\bbE\l[\int_t^se^{-ru}\partial_xBS_u\mu du\Big|X_t,\Sigma^2_t\r] \nonumber \\
                                 &\hspace{5mm}+\bbE\l[\int_t^se^{-ru}\partial_{\sigma^2}BS_u(-\lambda\Sigma^2_u)du\Big|X_t,\Sigma^2_t\r] \nonumber \\
                                 &\hspace{5mm}+\bbE\l[\int_t^se^{-ru}\int_0^\infty\Delta^{\rho z,z}BS_u\nu(dz)du\Big|X_t,\Sigma^2_t\r].
\end{align}
Taking the limitation on the left hand side as $s$ tends to $T$, we have
\[
\lim_{s\to T}\bbE[BS_s|X_t,\Sigma^2_t]=\bbE[BS_T|X_t,\Sigma^2_t],
\]
since $|BS_s|\leq\sup_{t\in[0,T]}S_t+K$, which is integrable.
Next, the partial derivatives $\partial_xBS$ and $\partial_{\sigma^2}BS$ are positive by (\ref{eq-dBSx}) and (\ref{eq-G}).
Thus, the monotone convergence theorem provides that
\begin{equation}\label{eq-thm1-21}
\lim_{s\to T}\bbE\l[\int_t^se^{-ru}\partial_xBS_udu\Big|X_t,\Sigma^2_t\r] = \bbE\l[\int_t^Te^{-ru}\partial_xBS_udu\Big|X_t,\Sigma^2_t\r]
\end{equation}
and
\[
\lim_{s\to T}\bbE\l[\int_t^se^{-ru}\partial_{\sigma^2}BS_u(-\lambda\Sigma^2_u)du\Big|X_t,\Sigma^2_t\r]
= \bbE\l[\int_t^Te^{-ru}\partial_{\sigma^2}BS_u(-\lambda\Sigma^2_u)du\Big|X_t,\Sigma^2_t\r].
\]
Moreover, from the view of the proof of Lemma \ref{lem2}, the dominated convergence theorem implies
\begin{align*}
& \lim_{s\to T}\bbE\l[\int_t^se^{-ru}\int_0^\infty\Delta^{\rho z,z}BS_u\nu(dz)du\Big|X_t,\Sigma^2_t\r] \\
&\hspace{5mm}=\bbE\l[\int_t^Te^{-ru}\int_0^\infty\Delta^{\rho z,z}BS_u\nu(dz)du\Big|X_t,\Sigma^2_t\r].
\end{align*}
To summarize the above, taking the limitation on both sides of (\ref{eq-thm1-2}) as $s$ tends to $T$, and multiplying $e^{rt}$ on both sides, we obtain
\begin{align}\label{eq-thm1-3}
V_t &= BS_t+\bbE\l[\int_t^Te^{-r(u-t)}\partial_{\sigma^2}BS_u(-\lambda\Sigma^2_u)du\Big|X_t,\Sigma^2_t\r] \nonumber \\
    &  \hspace{5mm}+\bbE\l[\int_t^Te^{-r(u-t)}\int_0^\infty\l\{\Delta^{\rho z,z}BS_u+\partial_xBS_u(1-e^{\rho z})\r\}\nu(dz)du\Big|X_t,\Sigma^2_t\r] \nonumber \\
    &= BS_t+I_1+I_2+\bbE\l[\int_t^Te^{-r(u-t)}\olcalL BS_udu\Big|X_t,\Sigma^2_t\r],
\end{align}
since $\mu=\int_0^\infty(1-e^{\rho z})\nu(dz)$.

\noindent{\it Step 2.} \
We shall calculate the last term of (\ref{eq-thm1-3}).
First of all, we fix $t\in[0,T)$ arbitrarily, and define
\[
F(u,x,\sigma^2):=e^{-r(u-t)}\tau_u\olcalL BS(u,x,\sigma^2), \ \ \ u\in[t,T).
\]
Lemma \ref{lem3} ensures that, for any $s,t\in[0,T)$ with $t<s$, $\olcalL BS(u,x,\sigma^2)$ is a $C^{1,2,1}$-function on $[t,s]\times\bbR\times[e^{-\lambda T}\Sigma^2_0,\infty)$.
Remark that the domain of $\sigma^2$ is restricted to $[e^{-\lambda T}\Sigma^2_0,\infty)$ from the view of (\ref{sigma1}).
Ito's formula, together with (\ref{BS-PDE-calL}) in Lemma \ref{lem3}, implies
\begin{align}\label{eq-thm1-4}
F(s,X_s,\Sigma^2_s) &= F(t,X_t,\Sigma^2_t)-r\int_t^se^{-r(u-t)}\tau_u\olcalL BS_udu \nonumber \\
                    &\hspace{5mm}-\int_t^se^{-r(u-t)}\olcalL BS_udu+\int_t^se^{-r(u-t)}\tau_u\partial_t\olcalL BS_udu \nonumber \\
                    &\hspace{5mm}+\int_t^se^{-r(u-t)}\tau_u\partial_x\olcalL BS_u\l(r+\mu-\frac{\Sigma^2_u}{2}\r)du \nonumber \\
                    &\hspace{5mm}+\int_t^se^{-r(u-t)}\tau_u(\partial_x\olcalL BS_u)\Sigma_udW_u \nonumber \\
                    &\hspace{5mm}+\frac{1}{2}\int_t^Te^{-r(u-t)}\tau_u(\partial^2_x\olcalL BS_u)\Sigma^2_udu \nonumber \\
                    &\hspace{5mm}+\int_t^se^{-r(u-t)}\tau_u\partial_{\sigma^2}\olcalL BS_u(-\lambda\Sigma^2_u)du \nonumber \\
                    &\hspace{5mm}+\int_t^se^{-r(u-t)}\tau_u\int_0^\infty\Delta^{\rho z,z}\olcalL BS_{u-}N(du,dz) \nonumber \\
                    &= F(t,X_t,\Sigma^2_t)-\int_t^se^{-r(u-t)}\olcalL BS_udu+\int_t^se^{-r(u-t)}\tau_u\partial_x\olcalL BS_u\mu du \nonumber \\
                    &\hspace{5mm}+\int_t^se^{-r(u-t)}\tau_u(\partial_x\olcalL BS_u)\Sigma_udW_u
                                 +\int_t^se^{-r(u-t)}\tau_u\partial_{\sigma^2}\olcalL BS_u(-\lambda\Sigma^2_u)du \nonumber \\
                    &\hspace{5mm}+\int_t^se^{-r(u-t)}\tau_u\int_0^\infty\Delta^{\rho z,z}\olcalL BS_{u-}N(du,dz).
\end{align}
Remark that the above integral with respect to $N(du,dz)$ is also well-defined by Lemma \ref{lem5}.
Taking the conditional expectation on both sides of (\ref{eq-thm1-4}), we have
\begin{align}\label{eq-thm1-5}
F(s,X_s,\Sigma^2_s) &= F(t,X_t,\Sigma^2_t)-\bbE\l[\int_t^se^{-r(u-t)}\olcalL BS_udu\Big|X_t,\Sigma^2_t\r] \nonumber \\
                    &\hspace{5mm}+\bbE\l[\int_t^se^{-r(u-t)}\tau_u\partial_x\olcalL BS_u\mu du\Big|X_t,\Sigma^2_t\r] \nonumber \\
                    &\hspace{5mm}+\bbE\l[\int_t^se^{-r(u-t)}\tau_u\partial_{\sigma^2}\olcalL BS_u(-\lambda\Sigma^2_u)du\Big|X_t,\Sigma^2_t\r] \nonumber \\
                    &\hspace{5mm}+\bbE\l[\int_t^se^{-r(u-t)}\tau_u\int_0^\infty\Delta^{\rho z,z}\olcalL BS_u\nu(dz)du\Big|X_t,\Sigma^2_t\r]
\end{align}
by Lemmas \ref{lem4} and \ref{lem5}.

Now, we take limits as $s$ tends to $T$ on both sides of (\ref{eq-thm1-5}).
A similar argument with the proof of Lemma \ref{lem2} yields
\[
\lim_{s\to T}\bbE\l[\int_t^se^{-ru}\int_0^\infty\Delta^{\rho z,0}BS_u\nu(dz)du\Big|X_t,\Sigma^2_t\r]
= \bbE\l[\int_t^Te^{-ru}\int_0^\infty\Delta^{\rho z,0}BS_u\nu(dz)du\Big|X_t,\Sigma^2_t\r],
\]
from which, together with (\ref{eq-thm1-21}),
\[
\lim_{s\to T}\bbE\l[\int_t^se^{-r(u-t)}\olcalL BS_udu\Big|X_t,\Sigma^2_t\r] = \bbE\l[\int_t^Te^{-r(u-t)}\olcalL BS_udu\Big|X_t,\Sigma^2_t\r]
\]
holds.
In addition, we have
\[
\lim_{s\to T}\bbE\l[\int_t^se^{-r(u-t)}\tau_u\partial_x\olcalL BS_udu\Big|X_t,\Sigma^2_t\r]
=\bbE\l[\int_t^Te^{-r(u-t)}\tau_u\partial_x\olcalL BS_udu\Big|X_t,\Sigma^2_t\r],
\]
and
\begin{align*}
& \lim_{s\to T}\bbE\l[\int_t^se^{-r(u-t)}\tau_u\int_0^\infty\Delta^{\rho z,z}\olcalL BS_u\nu(dz)du\Big|X_t,\Sigma^2_t\r] \\
&= \bbE\l[\int_t^Te^{-r(u-t)}\tau_u\int_0^\infty\Delta^{\rho z,z}\olcalL BS_u\nu(dz)du\Big|X_t,\Sigma^2_t\r]
\end{align*}
from the views of the proofs of Lemmas \ref{lem4} and \ref{lem5}, respectively.
Summarizing the above with Lemmas \ref{lem6} and \ref{lem7}, we obtain
\begin{align*}
\bbE\l[\int_t^Te^{-r(u-t)}\olcalL BS_udu\Big|X_t,\Sigma^2_t\r]
&= F(t,X_t,\Sigma^2_t)+\bbE\l[\int_t^Te^{-r(u-t)}\tau_u\partial_x\olcalL BS_u\mu du\Big|X_t,\Sigma^2_t\r] \\
&\hspace{5mm}+\bbE\l[\int_t^Te^{-r(u-t)}\tau_u\partial_{\sigma^2}\olcalL BS_u(-\lambda\Sigma^2_u)du\Big|X_t,\Sigma^2_t\r] \\
&\hspace{5mm}+\bbE\Bigg[\int_t^Te^{-r(u-t)}\tau_u\int_0^\infty\Delta^{\rho z,z}\olcalL BS_u\nu(dz)du\Big|X_t,\Sigma^2_t\Bigg].
\end{align*}
This completes the proof of Theorem \ref{thm1}.
\fin

\subsection{Lemmas}

\begin{lem}\label{lem1}
\begin{equation}\label{eq-lem1}
\bbE\l[\int_t^se^{-ru}(\partial_xBS_u)\Sigma_udW_u\Big|X_t,\Sigma^2_t\r]=0.
\end{equation}
\end{lem}

\proof
Since $\whS$ is a square integrable martingale, $\int_0^t\whS_u\Sigma_udW_u$ is also a square integrable martingale.
Thus, (\ref{eq-dBSx}) yields that
\[
\bbE\l[\int_0^Te^{-2ru}(\partial_xBS_u)^2\Sigma^2_udu\r]\leq\bbE\l[\int_0^T\whS^2_u\Sigma^2_udu\r]<\infty,
\]
which implies (\ref{eq-lem1}).
\fin

\begin{lem}\label{lem2}
The integral
\[
\int_t^se^{-ru}\int_0^\infty\Delta^{\rho z,z}BS_{u-}N(du,dz)
\]
is well-defined, and we have
\[
\bbE\l[\int_t^se^{-ru}\int_0^\infty\Delta^{\rho z,z}BS_{u-}N(du,dz)\Big|X_t,\Sigma_t\r]=\bbE\l[\int_t^se^{-ru}\int_0^\infty\Delta^{\rho z,z}BS_u\nu(dz)du\Big|X_t,\Sigma_t\r]
\]
for any $s,t\in[0,T)$ with $t<s$.
\end{lem}

\proof
From the view of Subsection 4.3.2 (p.231) of Applebaum \cite{Applebaum}, it suffices to see
\[
\int_0^T\int_0^\infty\bbE[|\Delta^{\rho z,z}BS_u|]\nu(dz)du<\infty.
\]
Here, $C$ denotes a positive constant, which may vary from line to line.
For $d^\pm$ and $d^\pm_{\rho z,z}$ defined in (\ref{def-d1}) and (\ref{def-d3}) respectively, we have
\begin{align}\label{eq-lem2-2}
|d^\pm_{\rho z,z}-d^\pm|
&\leq \frac{|x-\log K+r\tau_t|}{\sqrt{\tau_t}}\l|\frac{1}{\sigma_z}-\frac{1}{\sigma}\r|+\frac{|\rho|z}{\sigma_z\sqrt{\tau_t}}+\frac{|\sigma_z-\sigma|\sqrt{\tau_t}}{2} \nonumber \\
&\leq \frac{|x-\log K+r\tau_t|}{\sqrt{\tau_t}}\frac{|\sigma-\sigma_z|}{\sigma\sigma_z}+\frac{|\rho|z}{\sigma\sqrt{\tau_t}}+\frac{z\sqrt{\tau_t}}{2(\sigma_z+\sigma)} \nonumber \\
&\leq \frac{|x-\log K+r\tau_t|}{\sqrt{\tau_t}}\frac{z}{2\sigma^3}+\frac{|\rho|z}{\sigma\sqrt{\tau_t}}+\frac{z\sqrt{\tau_t}}{4\sigma} \nonumber \\
&\leq C\l(\frac{|x|+1}{\sqrt{\tau_t}}+\sqrt{\tau_t}\r)\frac{z}{\sigma\wedge\sigma^3},
\end{align}
where $\sigma_z$ is defined in (\ref{eq-BS1}).
Now, (\ref{eq-lem2-2}) implies
\begin{align*}
\lefteqn{|\Delta^{\rho z,z}BS(t,x,\sigma^2)|} \\
 &=    |e^{x_z}\Phi(d^+_{\rho z,z})-Ke^{-r\tau_t}\Phi(d^-_{\rho z,z})-e^x\Phi(d^+)+Ke^{-r\tau_t}\Phi(d^-)| \\
 &\leq e^{x_z}|\Phi(d^+_{\rho z,z})-\Phi(d^+)|+e^x|e^{\rho z}-1|\Phi(d^+)+Ke^{-r\tau_t}|\Phi(d^-_{\rho z,z})-\Phi(d^-)| \\
 &\leq e^x\frac{1}{\sqrt{2\pi}}|d^+_{\rho z,z}-d^+|+e^x|\rho|z+Ke^{-r\tau_t}\frac{1}{\sqrt{2\pi}}|d^-_{\rho z,z}-d^-| \\
 &<    C(e^x+1)\l(\frac{|x|+1}{\sqrt{\tau_t}}+\sqrt{\tau_t}\r)\frac{z}{\sigma\wedge\sigma^3}+e^x|\rho|z \\
 &<    C(e^x+1)(|x|+1)\l(\frac{1}{\sqrt{\tau_t}}+\sqrt{\tau_t}+1\r)\frac{z}{1\wedge\sigma\wedge\sigma^3}.
\end{align*}
Note that the second inequality is derived from
\[
|\Phi(d^+_{\rho z,z})-\Phi(d^+)| = \l|\int_{d^+}^{d^+_{\rho z,z}}\phi(\vt)d\vt\r| \leq \frac{|d^+_{\rho z,z}-d^+|}{\sqrt{2\pi}},
\]
where $\phi$ is the probability density function of the standard normal distribution.
Since the volatility process $\Sigma$ is bounded from below by (\ref{sigma1}), we have
\begin{align}\label{eq-lem2-3}
\lefteqn{\int_0^T\int_0^\infty\bbE[|\Delta^{\rho z,z}BS_u|]\nu(dz)du} \nonumber \\
&\leq C\int_0^T\l(\frac{1}{\sqrt{\tau_u}}+\sqrt{\tau_u}+1\r)du\int_0^\infty z\nu(dz)\sqrt{\bbE\l[\l(\sup_{t\in[0,T]}S_t+1\r)^2\r]\bbE\l[\l(\sup_{t\in[0,T]}|X_t|+1\r)^2\r]} \nonumber \\
&<    \infty
\end{align}
by (\ref{eq-S2}) and (\ref{eq-X2}), from which Lemma \ref{lem2} follows.
\fin

\begin{lem}\label{lem3}
For any $t$, $s\in[0,T)$ with $t<s$, and any partial derivative operator $\partial\in\{\partial_t,\partial_x,\partial^2_x,\partial_{\sigma^2}\}$,
$\partial\olcalL BS(u,x,\sigma^2)$ exists for $(u,x,\sigma^2)\in[t,s]\times\bbR\times[e^{-\lambda T}\Sigma^2_0,\infty)$, and we have
\begin{equation}\label{eq-lem3}
\partial\olcalL BS(u,x,\sigma^2)=\olcalL\partial BS(u,x,\sigma^2).
\end{equation}
In particular,
\begin{equation}\label{BS-PDE-calL}
\calD^{BS}\olcalL BS(u,x,\sigma^2)=0
\end{equation}
holds for $(u,x,\sigma^2)\in[t,s]\times\bbR\times[e^{-\lambda T}\Sigma^2_0,\infty)$.
\end{lem}

\proof
First of all, we show (\ref{eq-lem3}) for $\partial_x$.
By the definition of $\olcalL$, (\ref{def-BS}) and (\ref{eq-dBSx}), we have
\begin{align*}
\partial_x\olcalL BS(u,x,\sigma^2)
&= \partial_x\int_0^\infty\calL^zBS(u,x,\sigma^2)\nu(dz) \\
&= \partial_x\int_0^\infty\bigg\{e^{x_z}\Phi(d^+_{\rho z})-Ke^{-r\tau_u}\Phi(d^-_{\rho z})-e^x\Phi(d^+)+Ke^{-r\tau_u}\Phi(d^-) \\
&\hspace{5mm}+e^x\Phi(d^+)(1-e^{\rho z})\bigg\}\nu(dz) \\
&= \partial_x\int_0^\infty\bigg\{e^{x_z}(\Phi(d^+_{\rho z})-\Phi(d^+))-Ke^{-r\tau_u}(\Phi(d^-_{\rho z})-\Phi(d^-))\bigg\}\nu(dz) \\
&= e^x(1+\partial_x)\int_0^\infty e^{\rho z}(\Phi(d^+_{\rho z})-\Phi(d^+))\nu(dz) \\
&\hspace{5mm}-Ke^{-r\tau_u}\partial_x\int_0^\infty(\Phi(d^-_{\rho z})-\Phi(d^-))\nu(dz).
\end{align*}
Remark that $d^\pm$ and $d^\pm_{\rho z}$ appeared in this proof are defined in (\ref{def-d1}) and (\ref{def-d2}) respectively,
but time parameter $t$ is replaced with $u$.
Note that
\[
|\Phi(d^+_{\rho z})-\Phi(d^+)| \leq \frac{|d^+_{\rho z}-d^+|}{\sqrt{2\pi}} = \frac{1}{\sqrt{2\pi}}\frac{|\rho|z}{\sigma\sqrt{\tau_u}}.
\]
Thus, $|\Phi(d^+_{\rho z})-\Phi(d^+)|$ is integrable with respect to $\nu(dz)$.
Moreover, since $\phi^\prime$ is bounded, that is, there is a constant $C_{\phi^\prime}>0$ such that 
\begin{equation}\label{eq-lem3-1}
|\phi^\prime(d)|<C_{\phi^\prime}
\end{equation}
for any $d\in\bbR$, we have
\begin{align*}
|\partial_x(\Phi(d^+_{\rho z})-\Phi(d^+))|
&= |(\partial_xd^+_{\rho z})\phi(d^+_{\rho z})-(\partial_xd^+)\phi(d^+)| \\
&= \frac{1}{\sigma\sqrt{\tau_u}}|\phi(d^+_{\rho z})-\phi(d^+)| \leq \frac{1}{\sigma\sqrt{\tau_u}}\frac{C_{\phi^\prime}|\rho|z}{\sigma\sqrt{\tau_u}},
\end{align*}
which is also integrable with respect to $\nu(dz)$.
Similarly, we can see the integrability of $|\partial_x(\Phi(d^-_{\rho z})-\Phi(d^-))|$.
Thus, (\ref{eq-lem3}) holds when $\partial=\partial_x$ from the view of the dominated convergence theorem.

As for $\partial^2_x$, we have
\begin{align*}
\partial^2_x\olcalL BS(u,x,\sigma^2)
&= \partial_x\olcalL\partial_xBS(u,x,\sigma^2) \\
&= \partial_x\int_0^\infty\bigg\{\partial_xBS(u,x_z,\sigma^2)-\partial_xBS(u,x,\sigma^2)+\partial^2_xBS(u,x,\sigma^2)(1-e^{\rho z})\bigg\}\nu(dz) \\
&= \partial_x\int_0^\infty\bigg\{e^{x_z}\Phi(d^+_{\rho z})-e^x\Phi(d^+)+\l(e^x\Phi(d^+)+\frac{e^x}{\sigma\sqrt{\tau_u}}\phi(d^+)\r)(1-e^{\rho z})\bigg\}\nu(dz) \\
&= \partial_x\int_0^\infty\bigg\{e^{x_z}(\Phi(d^+_{\rho z})-\Phi(d^+))+\frac{e^x}{\sigma\sqrt{\tau_u}}\phi(d^+)(1-e^{\rho z})\bigg\}\nu(dz) \\
&= e^x(1+\partial_x)\int_0^\infty\bigg\{e^{\rho z}(\Phi(d^+_{\rho z})-\Phi(d^+))+\frac{1}{\sigma\sqrt{\tau_u}}\phi(d^+)(1-e^{\rho z})\bigg\}\nu(dz)
\end{align*}
by (\ref{eq-dBSx2}).
Thus, we can show (\ref{eq-lem3}) for $\partial^2_x$ by a similar argument with the case of $\partial_x$.
Similarly, (\ref{eq-lem3}) holds for $\partial_{\sigma^2}$, since (\ref{eq-G}), together with (\ref{eq-lem3-1}), implies that
\begin{align}\label{eq-lem3-2}
& \l|\partial_{\sigma^2}\calL^zBS(u,x,\sigma^2)\r| \nonumber \\
&=    \l|\frac{\sqrt{\tau_u}}{2\sigma}e^x\l(e^{\rho z}\phi(d^+_{\rho z})-\phi(d^+)\r)+\frac{\sqrt{\tau_u}}{2\sigma}e^x\l(\phi(d^+)+\partial_xd^+\phi^\prime(d^+)\r)(1-e^{\rho z})\r| \nonumber \\
&\leq \frac{\sqrt{\tau_u}}{2\sigma}e^x\l\{e^{\rho z}|\phi(d^+_{\rho z})-\phi(d^+)|+\frac{C_{\phi^\prime}}{\sigma\sqrt{\tau_u}}(1-e^{\rho z})\r\} \nonumber \\
&\leq \frac{C_{\phi^\prime}e^x}{2\sigma^2}(e^{\rho z}|\rho|z+1-e^{\rho z}) \leq \frac{C_{\phi^\prime}e^x}{2e^{-\lambda T}\Sigma^2_0}(e^{\rho z}|\rho|z+1-e^{\rho z}),
\end{align}
which is integrable with respect to $\nu(dz)$.
On the other hand, noting that
\[
\partial_td^\pm=\frac{x-\log K}{2\sigma\tau_u^{\frac{3}{2}}}-\frac{\eta^\pm}{2\sigma\sqrt{\tau_u}}
\]
for $u\in[t,s]\subset[0,T)$, where $\eta^\pm$ is defined in (\ref{eq-BS1}), we can see (\ref{eq-lem3}) for $\partial_t$ similarly.

Summarizing the above, together with (\ref{BS-PDE}), we have (\ref{BS-PDE-calL}).
\fin

\begin{lem}\label{lem4}
\[
\bbE\l[\int_t^se^{-r(u-t)}\tau_u(\partial_x\olcalL BS_u)\Sigma_udW_u\Big|X_t,\Sigma^2_t\r]=0
\]
for any $s,t\in[0,T)$ with $t<s$.
\end{lem}

\proof
We show this lemma by the same way as the proof of Lemma \ref{lem1}.
To this end, recall that
\begin{align*}
\partial_x\olcalL BS(u,x,\sigma^2) &= \olcalL\partial_xBS(u,x,\sigma^2) \\
&= e^x\int_0^\infty\l\{e^{\rho z}\l(\Phi(d^+_{\rho z})-\Phi(d^+)\r)+\frac{\phi(d^+)}{\sigma\sqrt{\tau_u}}(1-e^{\rho z})\r\}\nu(dz).
\end{align*}
Thus, we have
\[
\l|\partial_x\olcalL BS(u,x,\sigma^2)\r|^2\leq \frac{e^{2x}}{2\pi\sigma^2\tau_u}\l\{\int_0^\infty\l(e^{\rho z}|\rho|z+1-e^{\rho z}\r)\nu(dz)\r\}^2,
\]
which implies
\[
\bbE\l[\int_t^se^{-2r(u-t)}\tau_u^2(\partial_x\olcalL BS_u)^2\Sigma^2_udu\r]\leq Ce^{2rT}T\bbE\l[\int_t^s\whS^2_udu\r] \leq Ce^{2rT}T^2\bbE\l[\sup_{u\in[0,T]}|\whS_u|^2\r]<\infty
\]
for some $C>0$.
This completes the proof of Lemma \ref{lem4}.
\fin

\begin{lem}\label{lem5}
The integral
\[
\int_t^se^{-r(u-t)}\tau_u\int_0^\infty\Delta^{\rho z,z}\olcalL BS_{u-}N(du,dz)
\]
is well-defined, and we have
\begin{align*}
& \bbE\l[\int_t^se^{-r(u-t)}\tau_u\int_0^\infty\Delta^{\rho z,z}\olcalL BS_{u-}N(du,dz)\Big|X_t,\Sigma^2_t\r] \\
&\hspace{5mm}=\bbE\l[\int_t^se^{-r(u-t)}\tau_u\int_0^\infty\Delta^{\rho z,z}\olcalL BS_u\nu(dz)du\Big|X_t,\Sigma^2_t\r]
\end{align*}
for any $s,t\in[0,T)$ with $t<s$.
\end{lem}

\proof
By the same manner as Lemma \ref{lem2}, it suffices to see
\begin{equation}\label{eq-lem5-1}
\int_0^T\tau_u\int_0^\infty\bbE[|\Delta^{\rho z,z}\olcalL BS_u|]\nu(dz)du<\infty.
\end{equation}
Recall that
\begin{align*}
\olcalL BS(t,x,\sigma^2)
&= \int_0^\infty\bigg\{e^{x_z}\Phi(d^+_{\rho z})-Ke^{-r\tau_u}\Phi(d^-_{\rho z})-e^x\Phi(d^+)+Ke^{-r\tau_u}\Phi(d^-) \\
&\hspace{5mm}+e^x\Phi(d^+)(1-e^{\rho z})\bigg\}\nu(dz) \\
&= \int_0^\infty\bigg\{e^{x_z}\l(\Phi(d^+_{\rho z})-\Phi(d^+)\r)-Ke^{-r\tau_u}\l(\Phi(d^-_{\rho z})-\Phi(d^-)\r)\bigg\}\nu(dz).
\end{align*}
This implies
\begin{align}\label{eq-lem5-2}
& \Delta^{\rho z,z}\olcalL BS(t,x,\sigma^2) \nonumber \\
&= \int_0^\infty\bigg\{\calL^wBS(t,x_z,\sigma_z^2)-\calL^wBS(t,x,\sigma^2)\bigg\}\nu(dw) \nonumber \\
&= \int_0^\infty\bigg\{e^{x_{z+w}}\l(\Phi(d^+_{\rho z+\rho w,z})-\Phi(d^+_{\rho z,z})\r)-Ke^{-r\tau_u}\l(\Phi(d^-_{\rho z+\rho w,z})-\Phi(d^-_{\rho z,z})\r) \nonumber \\
&\hspace{5mm}-e^{x_w}\l(\Phi(d^+_{\rho w})-\Phi(d^+)\r)+Ke^{-r\tau_u}\l(\Phi(d^-_{\rho w})-\Phi(d^-)\r)\bigg\}\nu(dw) \nonumber \\
&= \int_0^\infty\bigg\{e^{x_{z+w}}\int_{d^+_{\rho z,z}}^{d^+_{\rho z+\rho w,z}}\phi(\vt)d\vt-Ke^{-r\tau_u}\int_{d^-_{\rho z,z}}^{d^-_{\rho z+\rho w,z}}\phi(\vt)d\vt \nonumber \\
&\hspace{5mm}-e^{x_w}\int_{d^+}^{d^+_{\rho w}}\phi(\vt)d\vt+Ke^{-r\tau_u}\int_{d^-}^{d^-_{\rho w}}\phi(\vt)d\vt\bigg\}\nu(dw) \nonumber \\
&= \frac{\rho}{\sigma_z\sqrt{\tau_t}}\int_0^\infty\int_0^w\bigg\{e^{x_{z+w}}\phi(d^+_{\rho z+\rho\zeta,z})-Ke^{-r\tau_u}\phi(d^-_{\rho z+\rho\zeta,z})\bigg\}d\zeta\nu(dw) \nonumber \\
&\hspace{5mm}-\frac{\rho}{\sigma\sqrt{\tau_t}}\int_0^\infty\int_0^w\bigg\{e^{x_w}\phi(d^+_{\rho\zeta})-Ke^{-r\tau_u}\phi(d^-_{\rho\zeta})\bigg\}d\zeta\nu(dw) \nonumber \\
&= \frac{\rho}{\sigma_z\sqrt{\tau_t}}\int_0^\infty\int_0^w\bigg\{e^{x_{z+w}}\phi(d^+_{\rho z+\rho\zeta,z})-e^{x_{z+\zeta}}\phi(d^+_{\rho z+\rho\zeta,z})\bigg\}d\zeta\nu(dw) \nonumber \\
&\hspace{5mm}-\frac{\rho}{\sigma\sqrt{\tau_t}}\int_0^\infty\int_0^w\bigg\{e^{x_w}\phi(d^+_{\rho\zeta})-e^{x_\zeta}\phi(d^+_{\rho\zeta})\bigg\}d\zeta\nu(dw) \nonumber \\
&= \frac{\rho e^x}{\sqrt{\tau_t}}\int_0^\infty\int_0^w(e^{\rho w}-e^{\rho\zeta})\bigg\{\frac{e^{\rho z}}{\sigma_z}\phi(d^+_{\rho z+\rho\zeta,z})
   -\frac{1}{\sigma}\phi(d^+_{\rho\zeta})\bigg\}d\zeta\nu(dw).
\end{align}
Note that the fifth equality of (\ref{eq-lem5-2}) comes from the following general fact:
\[
e^x\phi(d^+)=Ke^{-r\tau_t}\phi(d^-)
\]
for any $t\in[0,T)$, $x\in\bbR$ and $\sigma>0$.
In addition, the following inequality holds:
\begin{align*}
& \bigg|\frac{e^{\rho z}}{\sigma_z}\phi(d^+_{\rho z+\rho\zeta,z})-\frac{1}{\sigma}\phi(d^+_{\rho\zeta})\bigg| \\
&\leq \phi(d^+_{\rho z+\rho\zeta,z})\l|\frac{e^{\rho z}}{\sigma_z}-\frac{1}{\sigma}\r|+\frac{1}{\sigma}|\phi(d^+_{\rho z+\rho\zeta,z})-\phi(d^+_{\rho\zeta})| \\
&\leq \frac{1}{\sqrt{2\pi}}\l|\frac{e^{\rho z}-1}{\sigma_z}+\frac{1}{\sigma_z}-\frac{1}{\sigma}\r|
  +\frac{C_{\phi^\prime}}{\sigma}\bigg\{|d^+_{\rho z,z}-d^+|+\frac{|\rho|\zeta}{\sqrt{\tau_t}}\l|\frac{1}{\sigma_z}-\frac{1}{\sigma}\r|\bigg\} \\
&\leq \frac{1}{\sqrt{2\pi}}\l(\frac{|\rho|z}{\sigma}+\frac{z}{2\sigma^3}\r)+\frac{C_{\phi^\prime}}{\sigma}\bigg\{C\l(\frac{|x|+1}{\sqrt{\tau_t}}+\sqrt{\tau_t}\r)
      \frac{z}{\sigma\wedge\sigma^3}+\frac{|\rho|\zeta}{\sqrt{\tau_t}}\frac{z}{2\sigma^3}\bigg\}
\end{align*}
for some $C>0$.
Remark that $C_{\phi^\prime}$ is the positive constant defined in (\ref{eq-lem3-1}), and the last inequality is due to (\ref{eq-lem2-2}).
Thus, (\ref{eq-lem5-2}) is less than
\begin{align*}
& Ce^x(|x|+1)\l(\frac{1}{\tau_t}+\frac{1}{\sqrt{\tau_t}}+1\r)\frac{z}{\sigma\wedge\sigma^4}\int_0^\infty(w\wedge w^2)\nu(dw)
\end{align*}
for some $C>0$.
As a result, substituting $u$, $X_u$ and $\Sigma^2_u$ for $t$, $x$ and $\sigma^2$ respectively, we can see (\ref{eq-lem5-1}) by a similar way with (\ref{eq-lem2-3}).
\fin

\begin{lem}\label{lem6}
$\lim_{s\to T}F(s,x,\sigma^2)=0$ for any $x\in\bbR$ and $\sigma>0$.
\end{lem}

\proof
First of all, we have
\[
\tau_s\olcalL BS(s,x,\sigma^2) = \tau_s\int_0^\infty\Big\{e^{x_z}\l(\Phi(d^+_{\rho z})-\Phi(d^+)\r)-Ke^{-r\tau_s}\l(\Phi(d^-_{\rho z})-\Phi(d^-)\r)\Big\}\nu(dz).
\]
Now, we evaluate the above integrand as follows:
\begin{align*}
& \tau_s\Big|e^{x_z}\l(\Phi(d^+_{\rho z})-\Phi(d^+)\r)-Ke^{-r\tau_s}\l(\Phi(d^-_{\rho z})-\Phi(d^-)\r)\Big| \\
&\leq \tau_s\Big\{e^{x_z}\frac{|\rho|z}{\sqrt{2\pi}\sigma\sqrt{\tau_s}}+K\frac{|\rho|z}{\sqrt{2\pi}\sigma\sqrt{\tau_s}}\Big\}
      \leq \sqrt{T}\frac{|\rho|z}{\sqrt{2\pi}\sigma}(e^x+K),
\end{align*}
which is integrable with respect to $\nu(dz)$.
Thus, the dominated convergence theorem implies
\[
\lim_{s\to T}F(s,x,\sigma^2) = \int_0^\infty\lim_{s\to T}e^{-r(s-t)}\tau_s\calL^zBS(s,x,\sigma^2)\nu(dz) = 0.
\]
\fin

\begin{lem}\label{lem7}
\begin{align*}
& \lim_{s\to T}\bbE\l[\int_t^se^{-r(u-t)}\tau_u\partial_{\sigma^2}\olcalL BS_u(-\lambda\Sigma^2_u)du\Big|X_t,\Sigma^2_t\r] \\
&\hspace{5mm}=\bbE\l[\int_t^Te^{-r(u-t)}\tau_u\partial_{\sigma^2}\olcalL BS_u(-\lambda\Sigma^2_u)du\Big|X_t,\Sigma^2_t\r].
\end{align*}
\end{lem}

\proof
By (\ref{eq-lem3-2}), we have
\[
|\partial_{\sigma^2}\olcalL BS(u,x,\sigma^2)| \leq C\frac{e^x}{2\sigma^2}
\]
for some $C>0$.
Thus, we can find a constant $C>0$ such that
\[
\l|\int_t^se^{-r(u-t)}\tau_u\partial_{\sigma^2}\olcalL BS_u(-\lambda\Sigma^2_u)du\r| \leq CT^2\sup_{u\in[0,T]}S_u,
\]
which is integrable with respect to $\bbP$.
Hence, Lemma \ref{lem7} follows by the dominated convergence theorem.
\fin

%
%
\setcounter{equation}{0}
\section{Conclusions}
An Al\`os type decomposition formula for the vanilla call option for the BNS model has been derived by using Ito's formula twice.
Figure \ref{fig1} shows that the values of $V_0$ are away from the values of $BS(0,X_0,\Sigma^2_0)$.
This indicates that we need to develop an approximate option pricing formula by using our decomposition formula, but we leave it to future works.
Besides, such an approximation would enable us to develop an approximation of implied volatilities and a calibration method for model parameters.

\begin{center}
{\bf Acknowledgments}
\end{center}
Takuji Arai gratefully acknowledges the financial support of the MEXT Grant in Aid for Scientific Research (C) No.18K03422.


\end{document}